\documentclass{mn2e}
\usepackage{psfig}

\def\ltsima{$\; \buildrel < \over \sim \;$}
\def\lsim{\lower.5ex\hbox{\ltsima}}
\def\gtsima{$\; \buildrel > \over \sim \;$}
\def\gsim{\lower.5ex\hbox{\gtsima}}

\begin{document}

\title[Environment of GRB~021004]{Time resolved spectroscopy of 
GRB~021004 reveals a clumpy extended wind}

\author[Lazzati et al.]{Davide Lazzati$^1$,
Rosalba Perna$^1$, Joseph Flasher$^1$, Vikram V. Dwarkadas$^2$, 
Fabrizio Fiore$^3$
\\
$^1$ JILA, University of Colorado, 440 UCB, Boulder, CO 80309-0440, USA \\
$^2$ Dept. of Astronomy and Astrophysics, University of Chicago, 5640 
S Ellis Ave, AAC 010c, Chicago IL 60637 \\
$^3$ INAF - Osservatorio Astronomico di Roma, via Frascati 33, Roma, 
I-00040, Italy}

\maketitle

\begin{abstract}
High resolution spectroscopy of GRB~021004 revealed a wealth of
absorption lines from several intermediate ionization species. The
velocity structure of the absorber is complex and material with
velocity up to $\gsim3000$~km~s$^{-1}$ is observed.  Since only the
blueshifted component is observed, the absorber is very likely to be
material closely surrounding the gamma-ray burst.  We use a
time-dependent photoionization code to track the abundance of the ions
over time. Thanks to the presence of absorption from intermediate
ionization states at long times, we can estimate the location and
mass of the components of the absorber.  We interpret those
constraints within the hypernova scenario showing that the mass loss
rate of the progenitor must have been $\sim10^{-4} \,M_\odot$ per
year, suggestive of a very massive star. In addition, the wind
termination shock must lie at a distance of at least $100$~pc,
implying a low density environment. The velocity structure of the
absorber also requires clumping of the wind at those large distances.
\end{abstract}

\section{Introduction}
The high luminosity of gamma-ray burst afterglows makes them ideal
sources to probe their surrounding environment through absorption
lines in their powerlaw, generally featureless spectra.  In
particular, time variability (or lack thereof) in the equivalent width
of absorption lines can be used to constrain the size and density of
the absorbing region surrounding the burst (Perna \& Loeb 1998).

The association of GRBs with the collapse of massive stars (Stanek et
al. 2003; Hjorth et al. 2003) makes absorption studies particularly
useful as a new way to probe star-forming regions at intermediate and
high redshifts and/or the last hundreds of years of the progenitor
evolution. Lines associated with the material expelled by the star
allow one to probe the velocity structure and metal content of the
ejecta. The high quality and well sampled temporal evolution of the
spectroscopic data gathered for GRB~021004 (Moeller et al. 2002;
Matheson et al. 2003; Mirabal et al. 2003; Schaefer et al. 2003; Fiore
et al. 2005; Starling et al. 2005) allows us for the first time to
perform a detailed analysis.

GRB~021004 was detected by HETE II and an optical afterglow was
detected within 10 minutes with an R-band magnitude of 15.34 (Fox et
al. 2002).  Superimposed on the standard power-law decay, the optical
light curve shows several bumps (Holland et al. 2003), which can be
interpreted either as density fluctuations (Lazzati et al. 2002; Heyl
\& Perna 2003), continuous activity of the inner engine or
inhomogeneities in the fireball energy content (Nakar, Piran \& Granot
2003). High resolution spectroscopy of the optical afterglow was
reported by Fiore et al. (2005).  Low-to-intermediate resolution
spectroscopy was reported by Moeller et al. (2003), Mirabal et
al. (2003), Matheson et al. (2003), Schaefer et al. (2003) and
Starling et al. (2005).  These observations revealed a large set of
absorption lines spanning a velocity range of about $3000$~km~s$^{-1}$
toward the observer.  This is the only burst to date for which such
high-velocity lines have been detected.  The possibility of observing
such an absorption system due to chance alignment of different
absorbers was discussed and ruled out in previous work (see Schaefer
et al. 2003; Mirabal et al. 2003; Fiore et al. 2005). Interpreted as
absorption from the local GRB environment, the results of the
spectroscopic follow-up studies are controversial.  Starling et
al. (2005) conclude that the lines must be coming from a fossil
stellar wind with hydrogen enrichment from a companion.  Mirabal et
al. (2003) and Schaefer et al. (2003) attribute the lines to shells of
material present around the progenitor. They also allow for the
possibility of clumping.  Neither of those studies took adequately
into account the effect of the flash ionization of the burst photons
on the environment.

High resolution spectroscopy ($R\sim50000$) was obtained by Fiore et
al. (2005).  Several narrow components were resolved down to a width
of a few tens of km~s$^{-1}$, showing that some of the components
observed at low resolution are in fact due to blending of finer
velocity structures.  Fiore et al. (2005) derived physical parameters
for the absorbing medium modeling the ionization ratios with the code
CLOUDY, which assumes ionization equilibrium. They obtained fairly
constant ionization parameters for the different absorbers. They
concluded that the absorber is stratified as a wind, since a
$n\propto{}r^{-2}$ profile gives a constant ionization parameter over
a broad range of distances.  However, due to the highly time-dependent
nature of the ionizing flux and the long recombination times with
respect to the ionization times, their conclusions should be taken with
caution.  In this paper we combine all the available spectroscopic
data and adopt a time-dependent photoionization code (Perna \& Lazzati
2002) to determine the physical conditions of the absorber.

This paper is organized as follows: in $\S$2 we discuss the spectra of
GRB~021004 and show that there is moderate evidence of
time-variability in the equivalent width (EW) the observed CIV line.
In $\S$3 we present our model for the light curve of GRB~021004 while
the code and the data modeling technique are discussed in $\S$4
together with the results of the fit.  We finally discuss our findings
in $\S$5.

\section{Spectrum of 021004}

\begin{figure}
\psfig{file=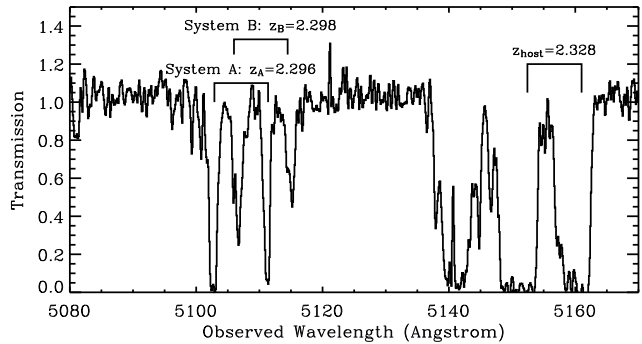,width=\columnwidth}
\caption{{A portion of the high resolution spectrum of GRB~021004
(Fiore et al. 2005) showing the complex absorption from CIV ions. The
two high velocity systems $A$ and $B$ discussed in the text are
shown.}
\label{fig:spettro}}
\end{figure}

\begin{figure}
\psfig{file=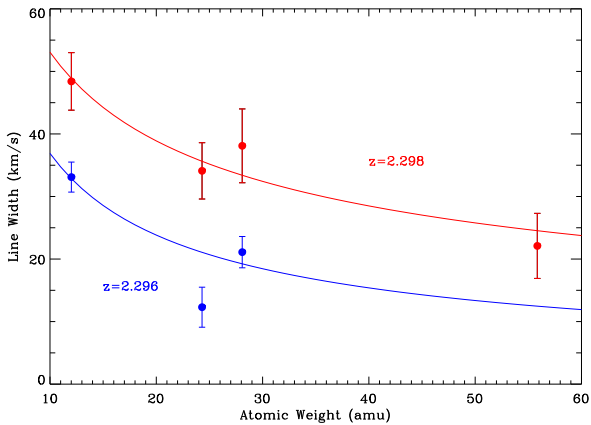,width=\columnwidth}
\caption{{Line widths vs. ion mass for absorption lines in the $A$ and
$B$ systems. Lines show fits with a thermal broadening model, which is
statistically consistent with the data.}
\label{fig:thermal}}
\end{figure}

Fiore et al. (2005) were able to identify six absorption systems in
the spectrum of GRB~021004 (see Fig.~1 of Fiore et al. 2005).  These
are characterized by the following velocities with respect to the
Ly$\alpha$ emission frequency which sets the reference frame of the
host galaxy ($z=2.328$): $v=0$~km~s$^{-1}$ ($z=2.328$);
$v\simeq-140$~km~s$^{-1}$ ($z=2.328$); $v\simeq-220$~km~s$^{-1}$
($z=2.328$); $v=\simeq-630$~km~s$^{-1}$ ($z=2.321$);
$v\simeq-2730$~km~s$^{-1}$ ($z=2.298$); $v\simeq-2900$~km~s$^{-1}$
($z=2.296$). The four low velocity systems are characterized by higher
column densities and severely saturated line cores which prevent a
robust measurement of the column density of the relative ions. They
can be explained in the framework of a wind environment as absorption
from material in the unshocked and shocked molecular cloud and/or from
a previous stage of mass ejection from the progenitor star (Van Marle
et al. 2005). We here concentrate on the two high velocity absorbers,
characterized by velocities of order $3000$~km~s$^{-1}$ that we call
system $A$ and system $B$ (see Fig.~\ref{fig:spettro}). The analysis
of the line widths of ions with different masses also shows that these
two systems are simple, their width being due to thermal motions of
the ions rather than to an underlying fine velocity structure
(Fig.~\ref{fig:thermal}). The derived temperatures are: $T=1 \times
10^6$~K ($z=2.298$); $T=3 \times 10^6$~K ($z=2.296$).  Throughout this
paper we focus on the absorption of the CIV doublet $1550+1548$~\AA\ and
of the SiIV line\footnote{Note that also this line is part of a
doublet but the companion line at 1404~\AA\ is blended with a lower
velocity component.} at $1393$~\AA.

\begin{figure}
\psfig{file=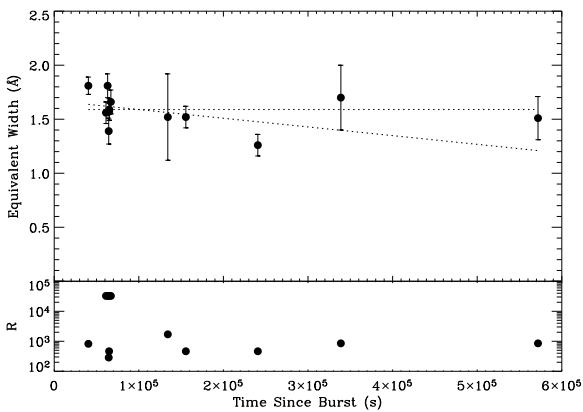,width=\columnwidth}
\caption{{The equivalent width of the CIV doublet taken from the
literature vs. time of observation. The best fit constant and
time-variable models are overlaid (see text for more details). The
F-test shows that the time-variable model provides a better
description of the data at the $90\%$ confidence.  The lower panel
shows the resolutions of the spectra from which the EW measurement has
been taken. They appear to be uncorrelated with time.}
\label{fig:ftest}}
\end{figure}

In order to explore possible variations of the equivalent widths of
these lines with time, we compiled all the data available in the
literature (see above for references). The CIV data (which have larger
signal to noise ratio than SiIV) are shown in Fig. \ref{fig:ftest}. A fit
with a constant yields $EW(t)=EW_0=1.59 \pm 0.03$ with a
$\chi^2/d.o.f.=26.6/12$.  To test for variability we model the EW
evolution with a linear function $EW(t)=EW_0+EW_1\,t$. We obtain
$EW_0=1.67 \pm 0.05$ and $EW_1=(-8.05 \pm 3.20) \times10^{-7}$ with
$\chi^2/d.o.f.=20.3/11$. An F-test shows moderate evidence of
variability at the $90\%$ level. Data have been taken with different
telescopes and spectral resolutions. To test weather this may cause a
spurious evolution of the EW, we show in the bottom panel of
Fig.~\ref{fig:ftest} the resolution with which each spectrum was
taken. The absence of any clear trend makes us more confident of the
reality of the detection of evolution in the CIV EW.

\section{Models}

\begin{figure*}
\parbox{0.97\columnwidth}{
\psfig{file=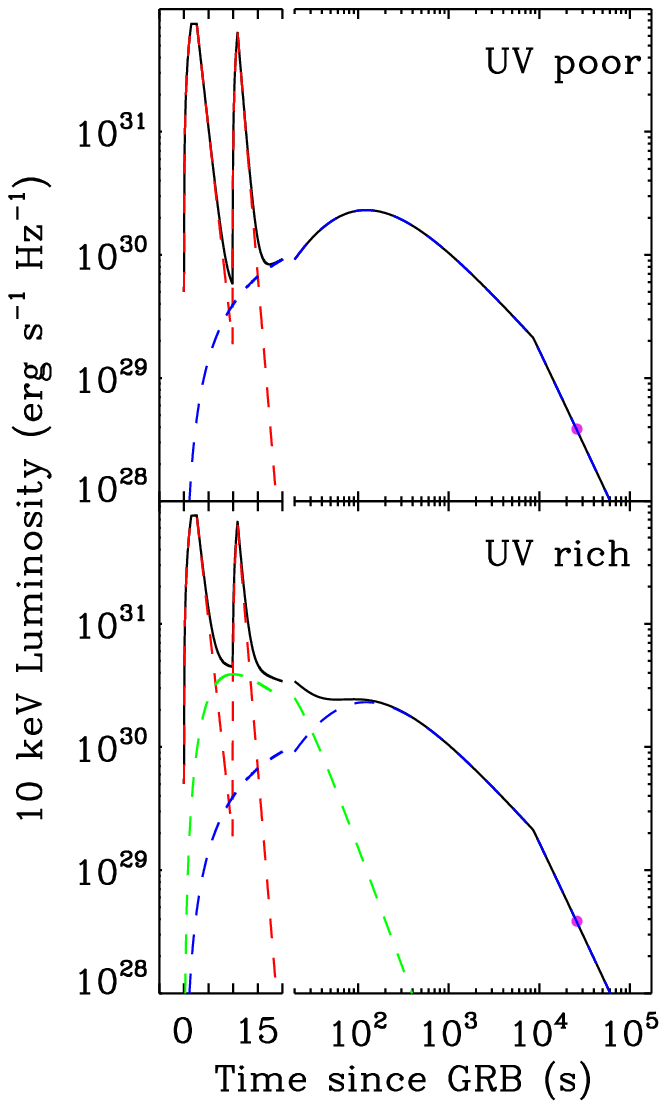,width=0.8\columnwidth}}
\parbox{0.97\columnwidth}{
\psfig{file=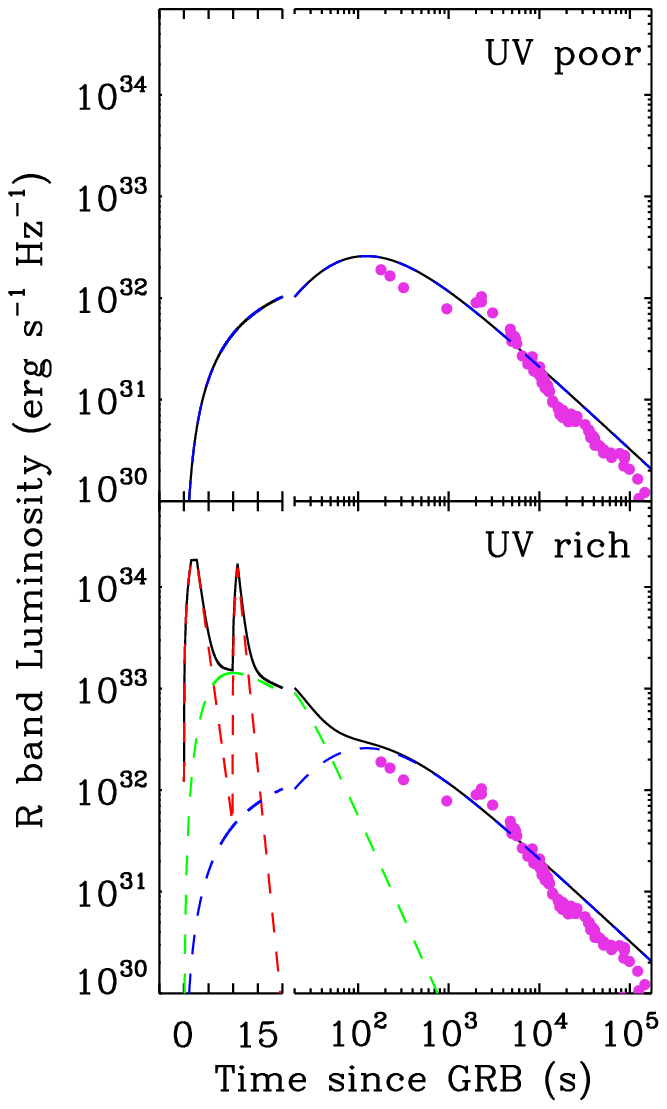,width=0.8\columnwidth}}
\caption{{{\em Left}: X-ray (10~keV observed) light curve of
GRB~021004 and its afterglow according to the model described in the
text. Note that the first 20 seconds of the time scale are linear
while the final ones are in logarithm. The top panels show the model
with less UV radiation in the prompt and optical flash phases among
those explored. The lower panels show the other extreme: a model in
which the UV component is emphasized as much as the optical and X-ray
constraints allow. The UV fluence is in all models dominated by the
afterglow component and therefore the results differ only marginally
between these two extremes.  {\em Right}: Same as in the left panel
but for the R-band.}
\label{fig:mod1}}
\end{figure*}

We model the light curve of GRB~021004 with three components: the
prompt phase, the afterglow and a possible optical flash.

The prompt phase is modeled with two pulses with a power-law spectrum
(Barraud et al. 2003):
\begin{equation}
L(\nu,t)=4\times10^{44}\,\nu^{-0.64}\,A(t) \qquad 
{\rm erg~s}^{-1}{\rm ~Hz}^{-1}
\label{eq:ldit}
\end{equation}
where $A(t)=A_1(t)+A_2(t)+A_3(t)+A_4(t)+A_5(t)$:
\begin{eqnarray}
A_1(t) &=& 0.157\,t\,\chi_{[0,1.5)} \nonumber \\
A_2(t) &=& 0.235 \, \chi_{[1.5,2.6)} \nonumber \\
A_3(t) &=& 0.235\,e^{-{{t-2.6}\over{1.22}}}\,\chi_{[2.6,\infty)} \\
A_4(t) &=& (0.2\,t-1.98)\,\chi_{[9.9,10.9)} \nonumber \\
A_5(t) &=& 0.2\,e^{-{{t-10.9}\over{0.9}}}\,\chi_{[10.9,\infty)} \nonumber
\end{eqnarray}
where $\chi_{(a,b)}$ is the characteristic function of the interval
$(a,b)$ ($\chi_{(a,b)}=1$ if $a<x<b$ and 0 otherwise). The
normalization constant in Eq.~\ref{eq:ldit} is chosen in such a way
that $\int \,L\,dt\,d\nu = 2\times10^{52}$~erg in the $[7-400]$~keV
band (Lamb et al. 2002).  Given the lack of spectral information below
$7$~keV observed ($23.2$~keV comoving), we allow for the possibility
of a sharp break in the spectrum at this frequency, with
$L(\nu)\propto\nu^2$ below $23.2$~keV.

The afterglow component (AG) is modeled with a broken power-law both
for the temporal and spectral components. We model the afterglow
spectral evolution in the framework of a fireball expanding in a
homogeneous medium (Lazzati et al. 2002, but see Li \& Chevalier 2003)
with a power-law slope $p=2.1$ of the relativistic electron
distribution (e.g. Holland et al. 2003). We obtain:
\begin{equation}
L(\nu,t)=5\times10^{32} {{2\,t_0^\alpha}\over{\left(t^{0.3\alpha}+
t_0^{0.3\left(\alpha-\beta\right)}\,t^{0.3\beta}\right)^{1/0.3}}}\,F(\nu,t)
\end{equation}
where $t_0=43.2$~s, $\alpha=-2$, $\beta=0.825$ and 
\begin{equation}
F(\nu,t)=\left\{
\begin{array}{ll}
\left({{\nu}\over{\nu_0}}\right)^{-0.55} & \nu<\nu_B(t) \\
\left({{\nu_B}\over{\nu_0}}\right)^{0.5}
\left({{\nu}\over{\nu_0}}\right)^{-1.05} & \nu\ge\nu_B(t)
\end{array}\right.
\end{equation}
where
\begin{equation}
\nu_B(t)=2.5\times10^{17}\,t^{-3/2}
\end{equation}
and the time is in seconds.

In addition to the prompt and afterglow components, we allow for the
presence of an optical flash (OF), as observed in GRB~990123 (Akerlof
et al. 1999). In the case of GRB~021004, optical observations at such
early times were not performed and the only constraints we have on
reverse shock emission come from the fact that it could only
marginally contribute to the afterglow at $t\gsim600$~s (see also
Kobayashi and Zhang 2003).  We model the OF as (e.g. Draine \& Hao
2002):
\begin{equation}
L(\nu,t)=2.2\times10^{41}{{\left({{t}\over{t_0}}\right)^2}\over{
\left[1+\left({{t}\over{t_0}}\right)^2\right]^2}}
\left\{\begin{array}{ll} \nu^{-0.5} & \nu<\nu_B\\
\nu_B^{0.7}\,\nu^{-1.2} & \nu\ge\nu_B
\end{array}\right.
\end{equation}
where $\nu_B=8\times10^{17}$~Hz.

While the afterglow is well constrained by the data, there is the
possibility of a break in the prompt phase and there may be an optical
flash.  We therefore initially considered four different models.  The
two extreme versions, with the smallest flux in the optical-UV and
with the highest flux in the optical-UV, are shown and compared to the
observed photometry in Fig.~\ref{fig:mod1}. Even though the models in
the figure look quite different, the optical-UV fluence is very
similar since it is dominated by the late afterglow component, for
which the spectrum is well constrained by the photometric data.
Indeed, by running test fits, we confirmed that the characteristics of
the absorber depend only marginally on the adopted spectral
model. Therefore we concentrated the detailed analysis only on one
model, the UV-poor one (upper panels of Fig.~\ref{fig:mod1}).

\subsection{The absorber geometry}
We explore two different geometries for the spatial distribution of
the absorber. Given the high velocity of the absorber, it must be
located in the unshocked part of the wind of the progenitor
star. Therefore we first model our absorber as a smooth wind with
density profile $n(R)\propto R^{-2}$. This model is characterized by
two fitting parameters: the mass loss rate $\dot{M}$ in units of solar
masses per year and the wind termination shock $R_{\rm{sh}}$ in
parsec. The presence of structure in the velocity of the absorber
suggests that the wind is clumped. We explore this possible geometry
as follows. Given the fact that our computations are line-of-sight, we
consider a thin shell geometry (with $\Delta R=0.1R$).  This
geometry is relevant if the clumps completely dominate the mass
distribution. The model is characterized by two fitting parameters:
the shell radius $R$ and the hydrogen column density $N_H$. Since the
recombination times of free electrons onto ions are much longer than
the times considered in the computation, we are insensitive to the
density of free electrons. As a consequence, fixing a shell width does
not imply any loss of generality.

The composition of the absorber is either a pure gas or a Milky Way
like interstellar material, with a dust component. The metallicity can
be varied as a free parameter. The dust component is a mixture of
graphite and silicates (Mathis, Rumpl \& Nordsieck 1977) with a size
distribution
\begin{equation}
\frac{dn_{i}}{da} = A_{i}n_{H}a^{-\beta}
\label{eqn:dist}
\end{equation}
where $\beta=3.5$ (Mathis et al. 1977) in the range $a_{min} \simeq$
0.005 $\mu$m, $a_{max} \simeq$ 0.25 $\mu$m. Given the data quality we
did not deem it worth pursuing a deeper investigation allowing for
different dust distributions.

Since the initial temperature of the absorbing material is not known,
we explored a range of temperatures $10^3<T_0<10^5$~K. The upper limit
represents the photospheric temperature of a massive WR star. We
considered also lower temperatures since the wind cools adiabatically
as it moves away from the star. We find that the late time ion
populations are insensitive to the initial temperature $T_0$ for
$T_0\lsim10^4$~K (see Table~\ref{tab:res1}).

\section{Methods and results}

\begin{figure*}
\parbox{\columnwidth}{
\psfig{file=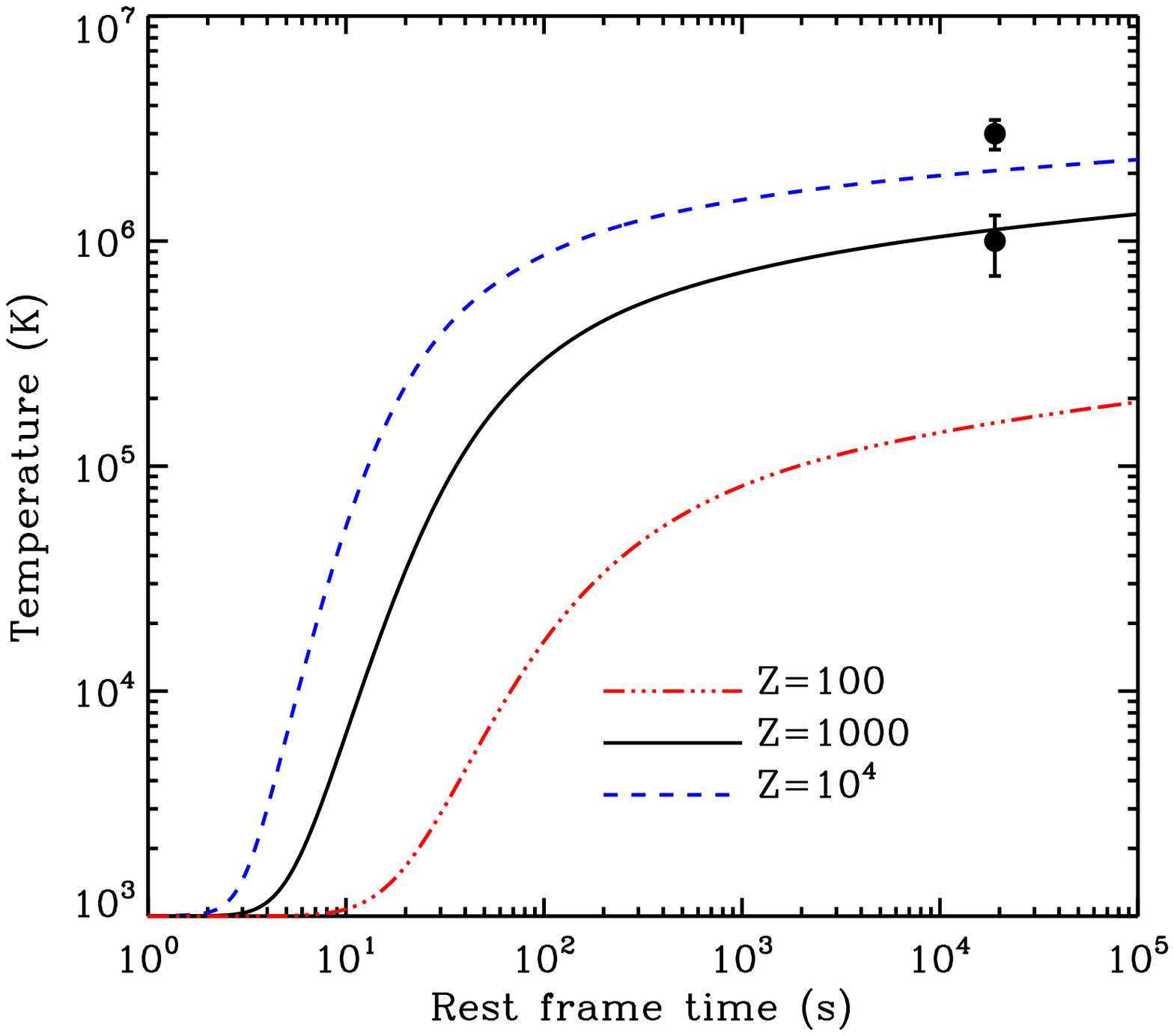,width=0.98\columnwidth}}
\parbox{\columnwidth}{
\psfig{file=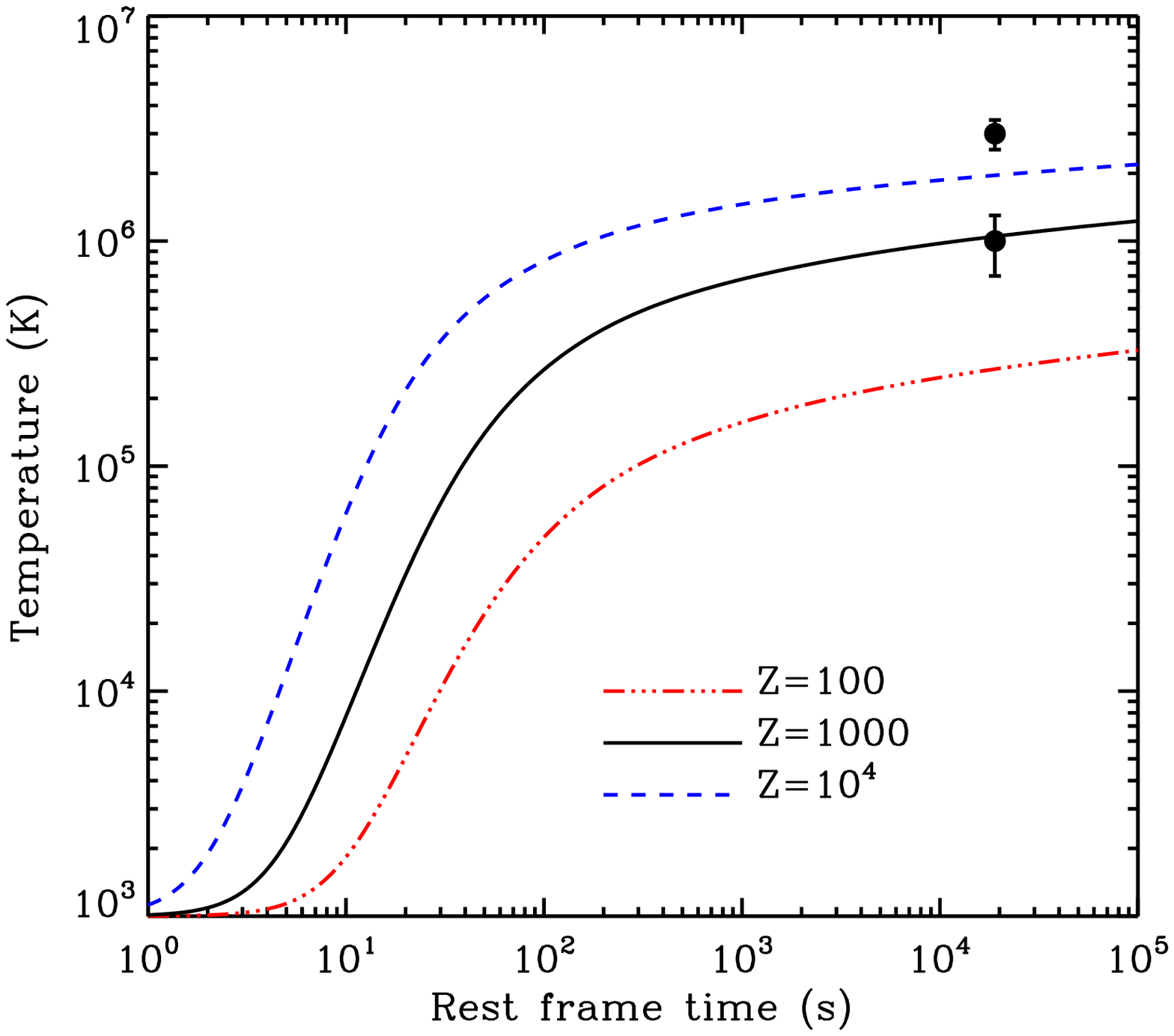,width=0.98\columnwidth}}
\caption{{Temperature of the absorbing material as a function of time
at a distance $R=100$~pc from the progenitor star. The left panel
shows the results for the wind geometry while the right panel shows
the shell geometry. Data are from the high resolution spectra of Fiore
et al. (2005) as modeled in Fig.~\ref{fig:thermal}.}\label{fig:temp}}
\end{figure*}

We use the time-dependent photoionization and dust-destruction code
by Perna \& Lazzati (2002; see also Perna, Lazzati \& Fiore 2003) to
follow the evolution of the dust distribution and of the ionization
state of the gas under the intense flux of burst and early afterglow
photons.

The column densities of CIV and SiIV provided by the simulations were
converted into EW by using the full Voigt profile:
\begin{equation}  
W_{\lambda} = \frac{\lambda^2}{c}\int^{\infty}_{0} 1 - 
e^{\frac{-\sqrt{\pi}e^2}{m_e c \Delta\nu_D} N f H(a,u)} d\nu
\label{eqn:ew}
\end{equation}
where $\Delta \nu_D$ is the Doppler width and $H(a,u)$ is the Voigt
function given by:
\begin{equation}
\Delta \nu_D = \frac{\nu_0}{c}\sqrt{\frac{2kT}{m}}
\label{eqn:dw}
\end{equation}
\begin{equation}
H(a,u) = \frac{a}{\pi} \int^{\infty}_{-\infty} \frac{e^{-y^2} 
dy}{a^2 + (u-y)^2}
\label{eqn:H}
\end{equation}
The Voigt function parameters are $a=\Gamma/(4\pi\Delta\,\nu_D)$ and
$u=(\nu-\nu_0)/(\Delta\,\nu_D)$ and $\Gamma$ is the transition rate,
approximated as $\Gamma=\gamma$, where $\gamma$ is the spontaneous
decay rate.

The two systems $A$ and $B$ that make the high velocity absorption
components are assumed to lie at the same distance from the burst
since the data are not accurate enough to perform an independent
fit. The two systems are blended in most of the medium resolution
spectra from the literature.  Thanks to the high resolution spectrum
of Fiore et al. (2005), we can measure the ratio of CIV and SiIV
column densities at $t=50$~ks as well as the temperature of the two
absorbers (see Fig.~\ref{fig:thermal}). We derive temperatures of
$T\sim3 \times 10^{6}$~K and $T\sim10^{6}$~K for the systems $A$ and
$B$, respectively. We also assume the same ionization for the two
systems and conclude that system $A$ contains three times more
material than system $B$.

The data for the high-velocity lines (the total EW for the $A$ plus
$B$ components) are given in Tab.~\ref{tab:highv}. There are two
additional model parameters that must be selected before fitting the
absorber properties as described above: the initial temperature of the
absorber and its metallicity. The only constraint on metallicity comes
from the spectroscopic observations of Mirabal et al. (2003) who
detect Ly$\alpha$ absorption with proper motion of
$\sim3000$~km~s$^{-1}$. Assuming that HI and CIV are the dominant
ionic species of hydrogen and carbon, respectively, a metallicity of
$\sim1000$ is derived. However, given the temperature measured above,
it is unlikely that the line ratio is a fair proxy to the metallicity
and this datum must be taken with caution.  For this reason we explore
metallicities between $100$ and $10^4$ solar, with particular emphasis
on the 1000 solar case (solar metallicities taken from Anders \&
Grevesse 1989). In all cases the He to H ratio is held solar as well
as the Fe and Ni to H ratios, since WR winds are not supposed to be
enriched of heavy elements that are synthesized in SN explosions.  As
we will see, the absorber temperature will provide additional evidence
for this value of the metallicity.

\begin{figure*}
\parbox{\columnwidth}{
\psfig{file=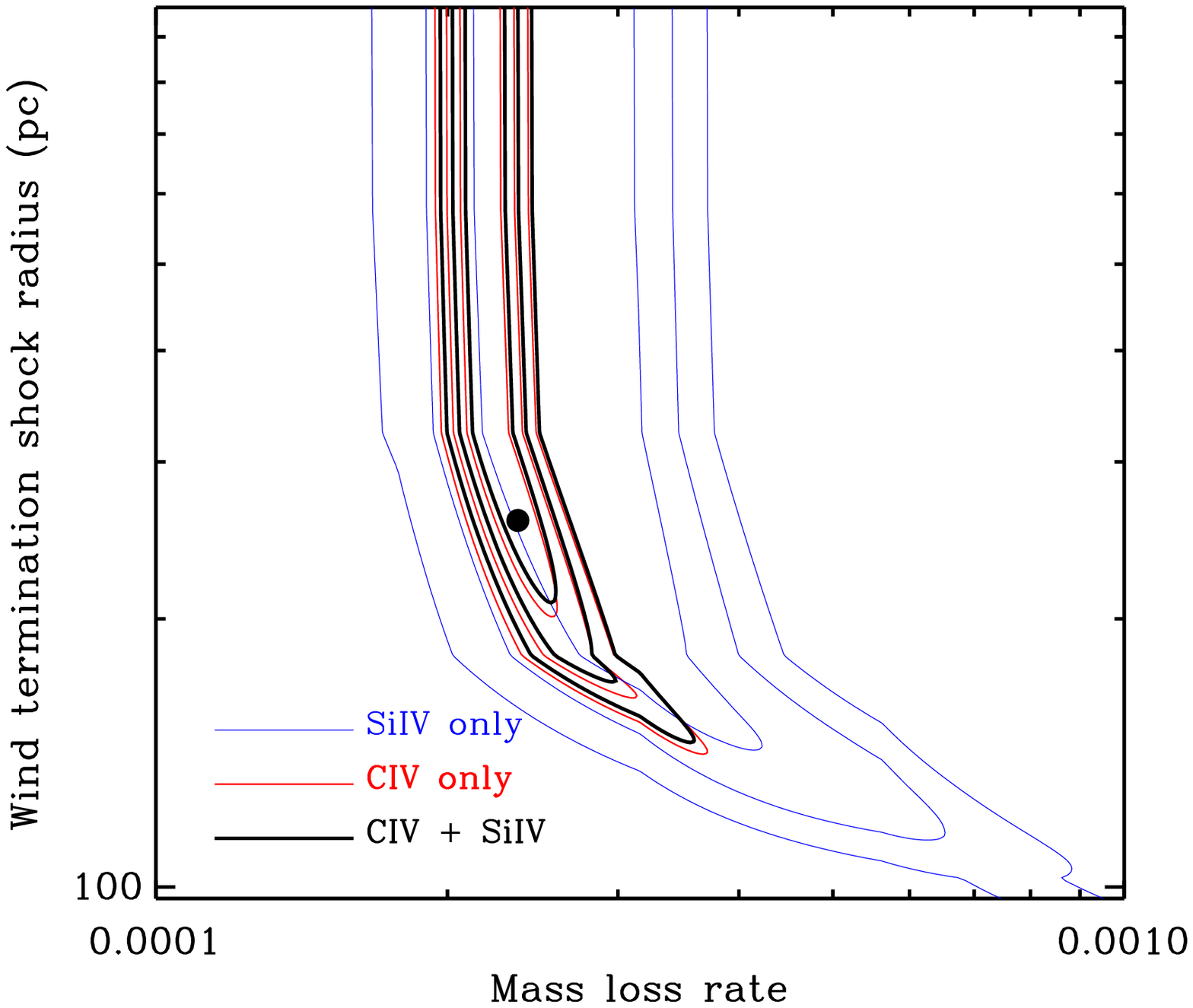,width=0.98\columnwidth}}
\parbox{\columnwidth}{
\psfig{file=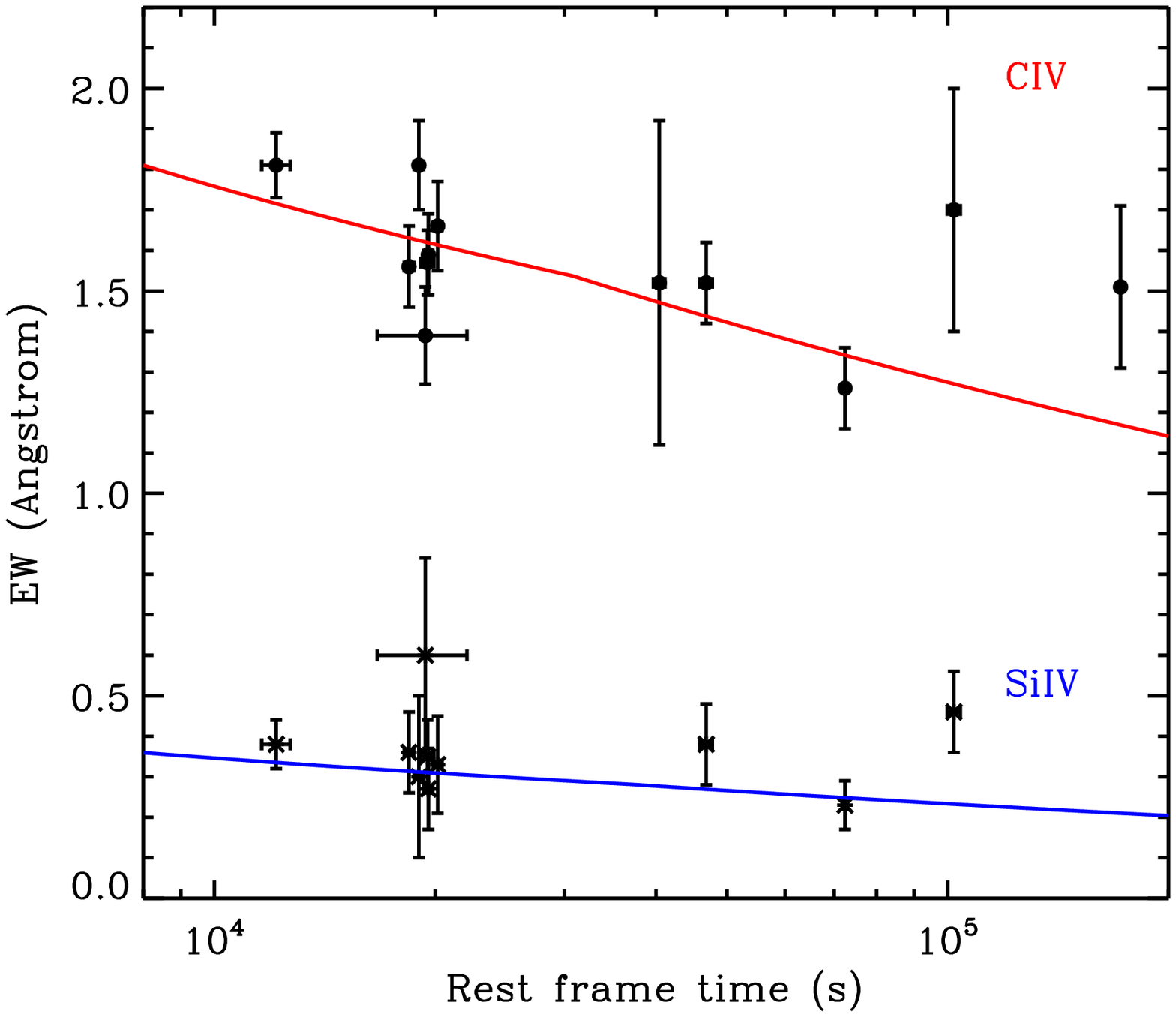,width=0.98\columnwidth}}
\caption{{Best fit models for the wind geometry case. The left panel
shows with a thick solid line the 1, 2 and 3 $\sigma$ confidence
regions of the best fit model for the $Z=1000\,Z_\odot$ metallicity
case. The broader contours (very thin blue line) show the confidence
contours if only the SiIV data are considered. The medium thickness
lines (red, almost indistinguishable from the thick ones) show the
contours if only CIV is considered. The right panel shows the data
with the best fit overlaid. }\label{fig:wind}}
\end{figure*}

The initial temperature of the absorber could in principle be
important since the ionic populations of CIV and SiIV vary sensibly
around $T\sim10^5$~K (Shull \& Van Steenberg 1982). In local thermal
equilibrium at $T=10^6$~K, only a minor fraction of CIV ($10^{-3.75}$)
is expected and no significant presence of SiIV. This implies that the
temperature of the absorbing blobs are affected by the burst
irradiation. The irradiation temperature of the absorber depends on
its metallicity, since the average ionization potential grows with
metallicity. In Figure~\ref{fig:temp} we show the temperature of the
absorber at the time of the high resolution spectroscopy of Fiore et
al. (2005). In both the wind and shell case, high metallicity
($1000\,Z_\odot<Z<10^4\,Z_\odot$) produces temperatures in the
observed range. A more accurate study cannot be performed since we fit
a single zone model for the two absorbers while they have different
temperatures. A different local metallicity or different radius can
explain the difference.

Starling et al. (2005) considered a GRB jet characterized by a double
structure, with a narrow component producing the GRB proper and a
wider outflow producing the afterglow. They argued that in this case a
lower mass loss rate and a smaller termination shock would agree with
the data since most of the absorbing material would avoid the flash
ionization caused by the prompt GRB phase.  To check this idea we ran
simulations with only the external shock afterglow component. We find
that these simulations do not differ from our UV-poor model (the one
widely explored in this work). Again, this is due to the fact that the
optical-UV fluence is dominated in all cases by the external shock
component. We therefore conclude that there is no evidence, from the
spectroscopic data, that the jet of GRB~021004 had multiple
components.

\begin{figure*}
\parbox{\columnwidth}{
\psfig{file=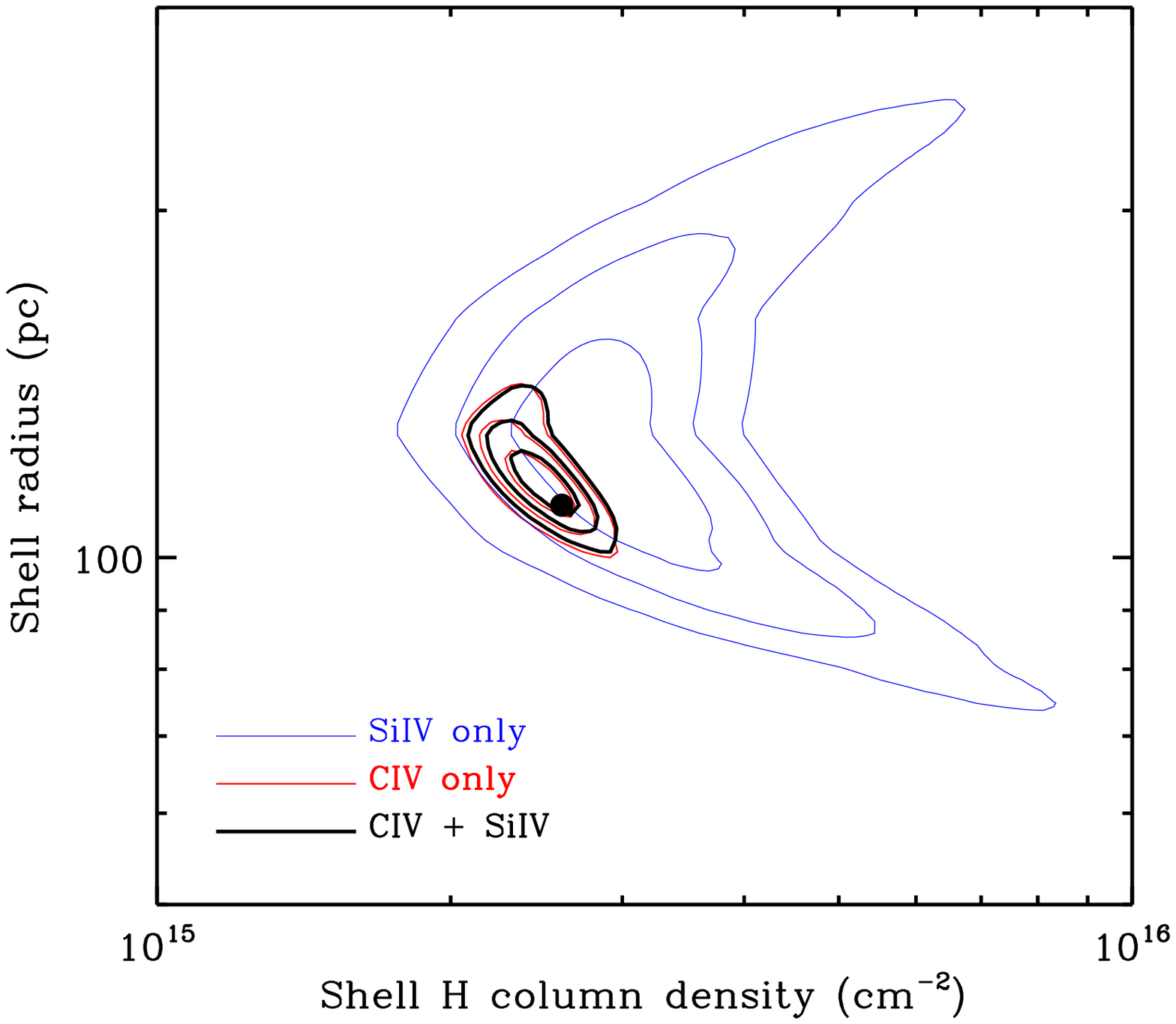,width=0.98\columnwidth}}
\parbox{\columnwidth}{
\psfig{file=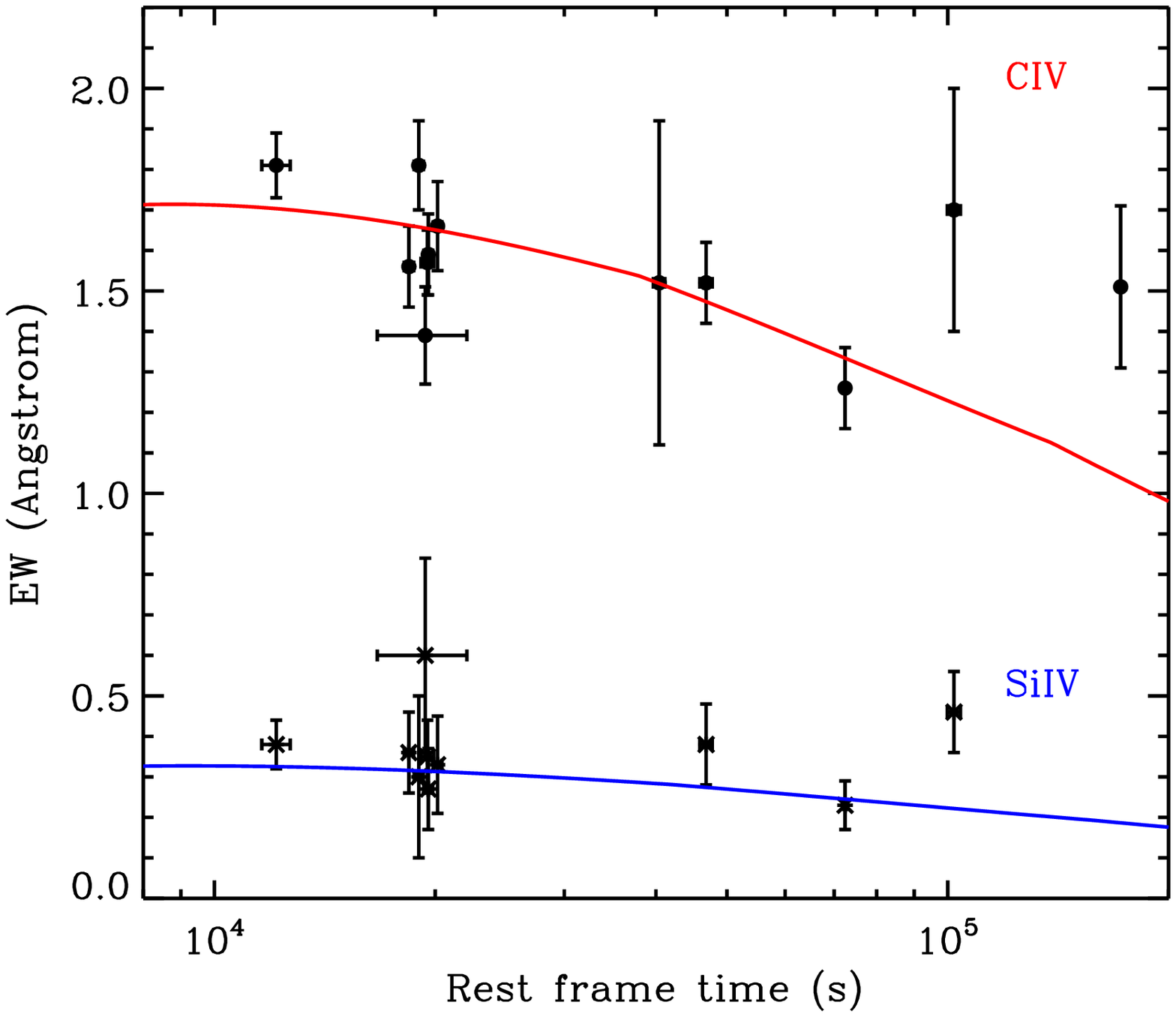,width=0.98\columnwidth}}
\caption{{Same as Figure~\ref{fig:wind} but for the shell case. 
}\label{fig:shell}}
\end{figure*}

\begin{figure}
\psfig{file=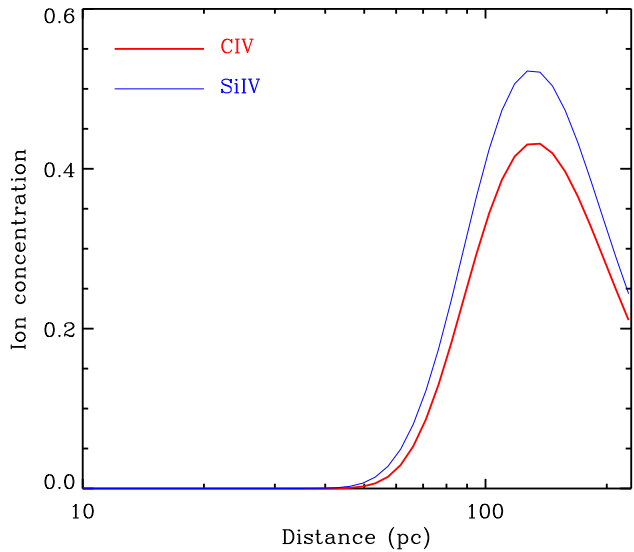,width=\columnwidth}
\caption{{The radial distribution of the concentration of CIV (thick
line) and SiIV (thin line) ions in the best fit wind model with
metallicity 1000 solar (see text for more details. The concentrations
are defined as the ratio of the ion density over the density of the
element in any ionization status.}\label{fig:winddistr}}
\end{figure}

The results of our fits are reported in Table~\ref{tab:res1} for the
wind case and in Table~\ref{tab:res2} for the shell (blob) case (see
also Figures~\ref{fig:wind} and \ref{fig:shell}). Statistically, the
two geometries are virtually indistinguishable. In both cases, the
absorber lies at a distance of about 100 parsecs from the progenitor
star (cf. also Fig.~\ref{fig:winddistr} for the wind case). This
result is at odds with what previously discussed in the literature
(Mirabal et al. 2003; Schaefer et al. 2003; Starling et al. 2005),
where the flash ionization of the GRB and the early afterglow was not
properly considered. It has important implications for the nature of
the progenitor star and the ISM in which it is embedded.

It is important to note that this lower limit on the distance to
the absorber is not due to the possible variability of the EWs. The
fact that some CIV and SiIV absorption is seen several days after the
GRB explosion requires the material to lie at a certain distance,
otherwise all the CIV would be ionized to CV or higher and all the
SiIV would be ionized to SiIV or higher. An order of magnitude
estimate can be obtained by considering that the afterglow of
GRB~021004 produced about $N_\gamma=4\times10^{60}$ CIV ionizing
photons. The ionization cross section of CIV is
$\sigma_{CIV}=7\times10^{-19}$~cm$^{2}$ (Verner \& Yakovlev 1995). It
follows that the distance of CIV ionization is approximately given, in
optically thin conditions, by:
\begin{equation}
R=\sqrt{\frac{N_\gamma\,\sigma_{CIV}}{4\pi}}\sim150 \;\;{\rm pc}\;.
\end{equation}
A more accurate estimate of the populations of CIV and SiIV can be seen
in Fig.~\ref{fig:winddistr}. 

In addition to the reported fit, we performed test runs for the other
continuum models described in $\S$~3. As anticipated, we find that the
change in the spectrum of the prompt emission affects the results only
at a marginal level. We also investigated the difference between a
dust rich (with solar dust-to-gas ratio and gas composition) and a
dust poor environment (with all the elements in gaseous phase). Since
the burst/afterglow are rich in UV radiation, the dust is thermally
sublimated out to large radii. For this reason the results with and
without an initial dust component are compatible within the errors, so
we do not report them in detail here.

For the wind case, the fit results are very similar, independent of
the metallicity. In all cases the mass loss rate is relatively high,
of the order of $10^{-4}$ solar masses per year. Given the distance
from the burster and the wind speed, the episode of high mass loss
took place approximately 30000 years before the GRB explosion.

\begin{table}
\begin{center}
\caption{Equivalent widths of the high-velocity lines. They are in Angstrom and in the GRB rest frame.}
\label{tab:highv}
\begin{tabular}{c|c|c|c|c|c|c}

Time&	error&	element&lines&		        	W$_{\lambda}$&	error&		ref.\\ \hline
40356.&	1800&	CIV&	1548+1550&			1.81&	.08&		(1)\\
40356.&	1800&	SiIV&	1393.76&			.38&	.06&		(1)\\
61202.7& 900&	CIV&	1548+1550&			1.56&	.1&		(3)\\
61202.7 &900&	SiIV&	1393.76	&			.36&	.1&		(3)\\
63127.2 &900&	CIV&	1548+1550&			1.81&	.11&		(4)\\
63127.2 &900&	SiIV&	1393.76	&			.3&	.2&		(4)\\
64422.	&9000&	CIV&	1548+1550&			1.39&	.12&		(5)\\
64422.	&9000&	SiIV&	1393.76	&			.6&	.24&		(5)\\
64872.0	&1300&	CIV&	1548+1550&			1.57&	.08&		(6)\\
64872.0	&1300&	SiIV&	1393.76	&			.35&	.09&		(6)\\
65044.1 &900 &    CIV&	1548+1550&			1.59&	.1&		(7)\\
65044.1 &900 &    SiIV&	1393.76	&			.27&	.1&		(7)\\
66968.5 &900&	CIV&	1548+1550&			1.66&	.11&		(8)\\
66968.5 &900&	SiIV&	1393.76		&		.33&	.12&		(8)\\
134275. &2400&	CIV&	1548+1550&			1.52&	.4&		(2)\\
155592. &3000&	CIV&	1548+1550&			1.52&	.1&		(6)\\
155592. &3000&	SiIV&	1393.76	&			.38&	.1&		(6)\\
240696. &900&	CIV&	1548+1550&			1.26&	.1&		(6)\\
240696. &900&	SiIV&	1393.76	&			.23&	.06&		(6)\\
338832.	&7000&	CIV&	1548+1550&			1.7&	.3&		(9)\\
338832.	&7000&	SiIV&	1393.76	&			.46&	.1&		(9)\\
571869. &3600&	CIV&	1548+1550&			1.51&	.2&		(2)

\end{tabular}
\end{center}
\footnotesize{References are: (1): Moeller et al. 2002; (2): Starling
  et al. (2005); (3): Our polarimetric spectra (0 rota); (4): Our
  polarimetric spectra (45 rota); (5): Schaefer et al. (2003); (6)
  Matheson et al. (2003); (7) Our polarimetric spectra (22 rota);(8)
  Our polarimetric spectra (67 rota); (9) Mirabal et al. (2003)}
\end{table}

\begin{table*}
\begin{center}
\caption{Fit results for the wind geometry cases.}
\label{tab:res2}
\begin{tabular}{c|c|c|c|c|c}
$T_0$ (K)      & $Z/Z_\odot$ & $R_{\rm{sh}}$ (pc) & $\dot{M}$ ($M_\odot$~y$^{-1}$) & $\chi ^{2}$ (prob)   \\ \hline
$10^3$         & $100$       & $>215$             & $(3.4\pm0.4)\times10^{-3}$     & $22.9$ ($30\%$)     \\
$10^3$         & $1000$      & $>210$             & $(2.3\pm0.3)\times10^{-4}$     & $22.9$ ($30\%$)   \\
$10^3$         & $10000$     & $>80$             & $(2.2\pm0.3)\times10^{-4}$     & $22.9$ ($30\%$)  
\end{tabular}
\end{center}
\end{table*}

\begin{table*}
\begin{center}
\caption{Fit results for the shell geometry cases.}
\label{tab:res1}
\begin{tabular}{c|c|c|c|c|c}
$T_0$ (K)      & $Z/Z_\odot$ & Radius (pc)       & $N_{H}$ ($10^{15}$ cm$^{-2}$)  & $\chi ^{2}$ (prob) & $M$ ($M_\odot$)   \\ \hline
$10^3$         & $1000$      & $116_{-6}^{+14}$  & $2.5\pm0.3$                    & $28.3$ ($11\%$)      & $86_{-3}^{+17}$   \\
$10^4$         & $1000$      & $116_{-5}^{+14}$  & $2.5\pm0.3$                    & $28.3$ ($11\%$)      & $86_{-4}^{+15}$   \\
$10^3$         & $100$       & $121_{-9}^{+14}$  & $28\pm3$                       & $26.5$ ($15\%$)    & $160_{-22}^{+44}$ \\
$10^3$         & $10000$     & $121_{-9}^{+14}$  & $0.28\pm0.03$                  & $26.8$ ($14\%$)    & $10_{-1.5}^{+2.5}$
\end{tabular}
\end{center}
\end{table*}

\section{Discussion}

We have collected all the available spectroscopic data of GRB~021004
from the literature and compiled the time resolved evolution of the EW
of CIV and SiIV absorption lines for the high velocity systems with
$v\sim3000$~km~s$^{-1}$. We have then used our time dependent
photoionization and dust-destruction code (Perna \& Lazzati 2002) to
model the evolution of the column densities of these ions under the
intense radiation field of the GRB and its early afterglow. Previous
work on subsets of these data (Mirabal et al. 2003; Schaefer et
al. 2003; Starling et al. 2005) concluded that the observed absorption
can be produced by material expelled by the GRB progenitor star under
the form of a smooth wind. Wolf Rayet stars are the principal candidate
due to the large speed and mass loss rate of their winds ((Nugis et
al.~1998; Willis 1999 ; Nugis \& Lamers 2000).

Wolf-Rayet stars are the final evolutionary stage of massive O-type
stars, with initial main-sequence masses above about 35 M$_{\odot}$
(for solar metallicities).  These stars eject material continuously in
the form of line-driven winds.  The observed mass loss rates of WC
stars (those WR stars that show strong carbon lines) are $\sim 1-6
\times 10^{-5}M_{\odot} {\rm yr}^{-1}$ and they produce winds with
terminal velocities 800 - 3500 km s$^{-1}$ (Nugis et al.~1998; Willis
1999 ; Nugis \& Lamers 2000).  In the case of GRB progenitors there
are not direct observations of the wind properties.  Rykoff et
al. (2004) argue that their mass loss may be substantially larger.
Fitting the afterglow photometric data of GRB~030418 they find that a
mass loss of $\sim10^{-3}\,M_\odot$~yr$^{-1}$ is required.

WR wind terminal velocities can explain the observed blueshifted
velocities we see within the high velocity system without invoking any
further radiative acceleration mechanisms due to the GRB itself.  In
the conventional model of a wind-blown bubble formed by mass-loss from
a massive star, the fast wind from a massive star expands into the
slower ambient medium, which could be a fossil wind from a previous
epoch, or even a molecular cloud. The fast wind sweeps up the medium
into a thin, dense, swept-up shell. The high pressure behind this
shell drives a reverse, or wind termination shock, into the freely
expanding wind (Weaver et al.~1977; Dwarkadas 2005). Although WR stars
form a late evolutionary stage of a massive star, which may undergo a
large amount of mass-loss prior to the WR stage, this general picture
provides a useful guide.

In this work we have shown how, by considering the flash ionization of
the burst and early afterglow photons, we were able to further
constrain the wind properties. In fact, the intense radiation flux
fully photoionizes the gas and destroys dust particles out to a large
radius. We have found that, in order for the remaining portion of the
wind material at large radii to provide the required column density,
the termination shock of the wind must lie at at least $\sim100$~pc
from the star and the mass loss rate from the progenitor star must
have been large, of the order of $10^{-4}\,M_\odot$ per year.  These
constraints are much more restrictive than what concluded before and
allow us to infer some important properties of the progenitor
star. Once again, we emphasize how this conclusion does not depend on
the possible decrease of the EW of the CIV line with time, but on the
simple and robust observation of the presence of high velocity CIV
several days after the burst explosion.

\begin{figure}
\psfig{file=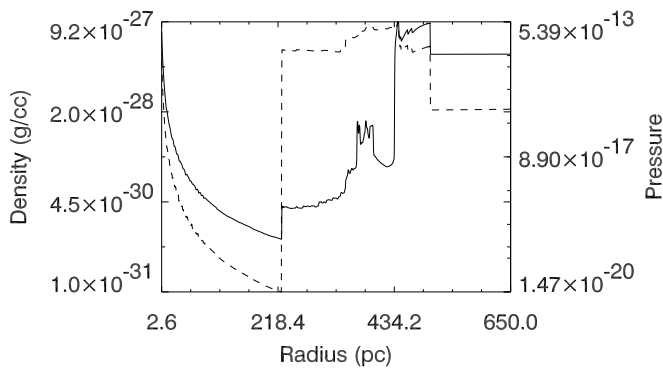,width=\columnwidth}
\caption{{Characteristics of the bubble formed by the wind of a star
of 120 $M_\odot$ expanding in a medium of density $10^{-3}$
cm$^{-3}$. The solid line shows the density (left y axis) while the
dashed line shows the pressure (right y axis).}
\label{fig:starmod}}
\end{figure}

The value of the termination shock that we have determined is in the
upper range of what generally discussed theoretically (Dwarkadas 2005;
Fryer, Rockefeller \& Young 2006). However, this large value can be
achieved by a very massive progenitor evolving in a low-density
environment. An illustrative example is shown in
Figure~\ref{fig:starmod}, where we used a stellar evolution model
provided by Meynet \& Maeder (2003) for a 120 solar mass star. The
model incorporates mass-loss from the star. We used the standard
assumption that the velocity of the emitted wind is a given multiple
of the escape velocity from the star at each time-step (Kudritzki \&
Puls 2000). The wind velocity and mass-loss rate are then used as
input parameters to compute the evolution of the surrounding medium by
means of the VH-1 code (Blondin \& Lundqvist 1993; Dwarkadas 2005).
As the star evolves, first as a main-sequence O star and then as a
Wolf-Rayet star, the wind mass-loss parameters gradually change. The
wind velocity, initially about 3500 km/s, drops to a low of about 1100
km/s and then increases again in the final stages to around 3000
km/s. The mass-loss rate increases from about $10^{-5} M_\odot$
yr$^{-1}$ to about twice that value. The ambient density was chosen to
be about 10$^{-3}$ particles~cm$^{-3}$, in order to have the wind
termination shock form at a large radius (about 215 pc in this
particular case).  Our inference of a low-density environment for the
progenitor of GRB~021004 could be interpreted within the context of
the recent observations that some GRB progenitors have been expelled
from their parent molecular clouds (Hammer et al. 2006). High velocity
in OB stars are indeed not uncommon (Mdzinarishvili \& Chargeishvili
2005). Another possibility for the creation of a low-density
environment is that the massive star progenitor of the GRB was in a
star cluster which created a superbubble; these structures are
commonly observed in our own Galaxy as well as in nearby ones
(e.g. Heiles 1978).

Finally, we have shown that a smooth wind is insufficient to fully
account for the observations (Mirabal et al. 2003; Schaefer et
al. 2003; Fiore et al. 2005; Starling et al. 2005).  The structure in
the velocity of the absorber suggests the presence of clumping, at
least at a moderate level.  To check if extreme clumping could explain
the observations, we modeled the absorber as material concentrated in
a shell of width $\Delta R=0.1R$, where $R$ is the radius of the
shell. This configuration is relevant for the case of clumps with an
angular size larger than the photon source (total covering\footnote{A
partial covering geometry is not allowed by the high-resolution data
that show line cores with no transparency.}). We found that the fits
are statistically equivalent, but the physical constraints on the mass
loss rate of the progenitor star are tighter
($\dot{M}\gsim10^{-3}\,M_{\odot}$ per year). This is due to the fact
that the material is more concentrated at a single radius with respect
to a smooth wind.

The high resolution spectra of Fiore et al. (2005) allow us to set
some constraints on the characteristics of the clumps.  Since the line
opacity is complete, the clumps must be larger than the source
size. In addition, since we see two clumps, the surface filling factor
must be close to unity.  At the time of the observations, the
transversal radius of the observable portion of the fireball is given
by:
\begin{equation}
R_{\rm{fb}}(t) = c\,t\,\Gamma = 3\times10^{16} \frac{t}{2{\rm{days}}} 
\frac{\Gamma}{5}
\qquad {\rm{cm}}
\end{equation}
This is likely to be smaller than the dimension of clumps in the wind
at a distance of $\sim100$ parsec from the star. 

It is possible that our estimate of the mass loss rate is moderately
affected by the wind inhomogeneity. Consider a clumpy wind with a
surface filling factor of $0.1$. In $90\%$ of the cases no high
velocity absorption would be seen.  Given that some absorption was
detected, in $10\%$ of the cases velocity structure would be
present. If this were the case, our $\dot{M}$ would be overestimated
by a factor $\sim10$.

These considerations on the presence of inhomogeneities in the wind 
of the GRB~021004  progenitor star should not come as a surprise. 
It has indeed been suggested for many years that WR winds are clumped (Moffat
et al. 1988; Li et al. 2000).  Clumping in the wind has been proposed
as one scenario to solve the so-called ``wind-momentum'' problem
present in these outflows.  It has been noted by several authors that
the momentum in the wind ${\dot{M}v_{\inf}}$ is several times the
momentum flux (L/c) from the star (Barlow et al. 1981). Although this
was initially explained away by scattering the radiation from the star
more than once, it is unclear if there is sufficient opacity in the
wind to accomplish this.  Clumps solve this problem in two ways 1) by
increasing the density and therefore increasing the luminosity by the
$n^2$ factor in the emissivity (but see Brown et al. 2004) and 2) the
consequence of clumping is to reduce empirically derived mass-loss
rates of hot stars by about a factor of 3, thus reducing the wind
momentum itself. 

In summary, the complex spectra of GRB~021004 can be explained as a
result of the absorption due to a very extended wind surrounding the
GRB progenitor star. It also requires a sizably large mass loss rate
of $\sim10^{-4}\,M_\odot$ per year. The large radial development of
the wind is likely due to a combination of the high mass loss, the
wind speed and of a low density environment. Some degree of clumping
is also required by the structure of the absorption in the velocity
space. This study shows how time resolved medium-high resolution
spectroscopy of the early GRB afterglow is a powerful tool to
investigate the properties of the GRB environment and constrain the
characteristics of the progenitor.

\section*{Acknowledgements}

This work was partially supported by NSF grants AST-0307502 (DL) and
AST~0507571 (JF, RP, DL) and NASA Astrophysics Theory Grant NAG5-12035
(DL) and NASA Swift Grant NNG05GH55G (JF, RP, DL). VVD's research is
supported by NSF grant AST 0319261. VVD is grateful to Joe Cassinelli
for stimulating email discussions on clumping in WR stars, and would
like to acknowledge interesting conversations with Don York and
Hsiao-Wen Chen on absorption lines in GRBs.

\end{document}